# Diagnostic Setup for Characterization of Near-Anode Processes in Hall Thrusters

L. Dorf, Y. Raitses and N. J. Fisch

Princeton Plasma Physics Laboratory (PPPL), Princeton, NJ, 08543

A diagnostic setup for characterization of near-anode processes in Hall-current plasma thrusters consisting of biased and emissive electrostatic probes, high-precision positioning system and low-noise electronic circuitry was developed and tested. Experimental results show that radial probe insertion does not cause perturbations to the discharge and therefore can be used for accurate near-anode measurements.

# I. INTRODUCTION

Over the past four decades comprehensive experimental studies were performed on Hall thrusters.[1,2,3] However, near-anode processes have not received sufficient experimental scrutiny. Recent theoretical models suggest that Hall thrusters may operate with or without a negative anode sheath.[4,5] The near-anode processes might then affect the overall operation of a Hall thruster. For example, a change of the voltage drop in the anode sheath might affect anode heating, propellant ionization,[6] or the beam divergence inside and outside the thruster.

Similar to the experimental studies of the sheath and presheath in low-pressure gas discharges,[7] the plasma parameters in the near-anode region can also be studied by various electrostatic probe techniques, including single, double and emissive probes. However, the implementation of conventional plasma diagnostics for measurements inside the Hall thruster channel is complicated by its relatively small size and obstacles imposed by thruster structures, in particular by the magnetic circuit. Stationary probes are limited to measurements near the outer channel wall. However, introducing movable probes axially into the acceleration region, with $T_e \sim 20$ eV, can cause significant perturbations to the Hall thruster discharge.[8,9,10] This might cause an inaccuracy of up to several volts in the measured plasma potential, which is of the order of the expected potential change over the entire near-anode region.

As we show in this paper, this significant disadvantage of movable probe diagnostics vanishes if the probe is introduced radially into the near-anode region and does not pass through the acceleration region. It does not cause significant perturbations to the discharge or get severely damaged, because the electron temperature and plasma density in the near-anode region are low. For the same reason, the probe residence time is not an issue. Therefore, radial probe insertion does not require building an expensive and complex high-speed positioning system.[11]

The near-anode region in Hall thrusters is typically about 1-2 cm long. Plasma density, $n \sim 10^{10} - 10^{11}$ cm$^{-3}$, electron temperature, $T_e \sim 3 - 5$ eV, and their variations in the near-anode region are smaller than in the acceleration region. The magnetic field is also much smaller, so the electron flux towards the anode is mainly affected by the electron pressure gradients, $\frac{1}{en}\frac{dP}{dz} \sim 10$ V/cm.

This paper is organized as follows. Technical questions specific for near-anode measurements in Hall thrusters are discussed in Section II. Section III describes the diagnostic setup built for the characterization of a near-anode region inside a 2 kW Hall thruster. In Section IV, we present testing results for biased and emissive probes.

## II. NEAR-ANODE MEASUREMENTS

The anode sheath thickness is typically assumed to be a few Debye lengths, where $\lambda_D \sim 0.05 mm$. Such a thin sheath does not easily accommodate probe diagnostics. The information about overall voltage drop in the sheath, however, can be obtained through probing plasma in the presheath, at several millimeters from the anode.

The absolute value of the sheath voltage drop is not expected to exceed 20V. The overall potential drop over the near-anode region is expected to be of the order of several volts. Therefore, uncertainties in interpreting the measured voltage are significantly decreased if plasma potential is measured relative to the anode rather than to the ground or the cathode. In Hall thrusters, the potential of the anode relative to the ground is several hundred volts. Therefore, anode-referenced measurements require the use of isolation amplifiers, which typically have DC-offsets. The calibration of amplifiers used in probe circuits showed that these offsets depend on the input load impedance, the discharge voltage, the mass flow rate and especially on the choice of reference potential.

The maximum input signal for commercial isolation amplifiers is typically 10 Volts, which necessitates the use of voltage dividers. For floating and plasma potential measurements with a floating emissive probe, the impedance of the divider, $R_{div}$, must be much larger than the impedance of the probe-to-plasma interface, $R_{sheath} = T_e /(e \cdot I_i^{sat})$ (ratio of electron temperature to ion saturation current and electron charge), to minimize the leakage current through the probe circuit. For certain Hall thruster operating regimes, $I_i^{sat}$ to the probe with collecting area of 3 mm$^2$ can be no more than several micro-amperes and therefore $R_{sheath}$ can reach 1 Mohm in the near-anode plasma, yielding $R_{div}$ must be hundreds of mega-ohms. In anode-referenced measurements, such a large load impedance causes the DC-offset of up to hundreds of millivolts, which can be of the order of overall changes of the measured signal. Therefore, for emissive probe measurements the ground appears to be a better reference than the anode, unlike for a single biased probe.

Another source of leakage current in a probe circuit can be the plasma collected by electrical connectors, which couple the probe wire, the extension cable and the electrical feed-through mounted on a flange of a vacuum vessel. Wrapping all connectors in Teflon tape and closing the face of the flange with the graphite foil eliminates this problem. This was shown by placing a test wire with the tip stripped from insulation and wrapped in Teflon tape inside the vacuum vessel. The wire was connected to the emissive probe circuitry (Fig. 2. b) and it measured zero floating potential at all thruster operating regimes.

## III. DIAGNOSTIC SETUP

The 2 kW Hall thruster and test facility used in this study are described in Ref. 11. Fig. 1 shows a diagnostic setup for near-anode measurements and the probe locations relative to the thruster channel. The probe holder, which can accommodate up to three plain electrostatic or hot emissive probes simultaneously, is mounted on a CVI precision rotary stage and a Newport linear stage for fine pitch and height adjustment of the probe relative to the thruster channel. These manually controlled stages are assembled on a Velmex motor-driven X-Y linear positioning stage, equipped with two 400-steps/rev step-motors, two 40-rev/inch high precision lead screws and two 5 µm-resolution Renishaw optical encoders. This motor-driven stage is mounted on an additional CVI rotary stage for a fine yaw adjustment, which is fixed on an aluminum breadboard near the thruster mounting table. Probes can be introduced into the thruster through a 2mm wide and 10 mm long axial slot starting at 2 mm from the anode, made in the outer ceramic wall of the thruster channel. The motor-driven positioning stage allows probe motion along the slot between the inner and the outer channel walls. Control of the positioning system and the signal measurements are performed by a National Instruments, PC-based, data acquisition system PCI-DIO-16-1.

The biased flat probe is constructed of 0.76 mm diameter thoriated tungsten rod, covered by a high purity alumina single bore tube with outer diameter of 1.3 mm and inner diameter of 0.79mm (Fig. 1). The probe collecting surface area, $A_{pr} = 0.45\ mm^2$, is much smaller than the anode cross-sectional area, $A_{an} = 7700\ mm^2$, so the collection of electrons by the probe, placed near the anode, does not affect the discharge. The overall probe length is 165 mm. The tungsten rod is coupled to the coaxial cable (silicon coated for vacuum compatibility) through the regular copper connector. The planar tip geometry was chosen because, in the near-anode region, the voltage-current characteristics of a flat probe appear to have more distinctive electron and ion saturation than those of a cylindrical probe. The probe sheath expansion and particle orbital motion are likely to account for this fact.[12]

The probe is biased relative to the anode with a KEPCO bipolar power supply BP200-1M, which is programmed with a one cycle sinusoidal signal by a PC-based function-generator connected through a Burr Brown isolation amplifier ISO124P (Fig. 2 a). A high-voltage vacuum switch is used to manually switch between 500 Ω and 100 KΩ shunts for measurements of electron and ion parts of the probe voltage-current characteristic, respectively. The probe current, $I_{PR}$, and biasing voltage, $V_B$, are measured through Analog Devices isolation amplifiers AD210AN.

The emissive probe is constructed of a 0.1 mm diameter thorriated tungsten wire covered by a 1.6 mm diameter and 82 mm long high-purity double-bore alumina tube. Each of the tube channels is filled with seven additional 0.1 mm diameter tungsten wires in order to prevent the heating of the probe aside from the tip. The probe tip protrudes from the alumina tube by 1.4 mm. The probe wires are coupled to a twisted shielded pair of 16 AWG wires (with a high-temperature Teflon insulation) through the regular copper connectors. A molybdenum tube extends the short alumina tube so that the overall probe length is 190 mm.

Similar to Ref. 13, the filament is heated with a 60 cycle half-wave rectified sinusoidal signal, as shown schematically in Fig 2.b. During the off half-cycles of the heating voltage, the voltage drop across the probe is assumed to be zero. An additional isolation transformer is placed in front of the variac for noise reduction. The floating potentials of the probe legs, $\Phi_{1,2}$, are measured relative to the ground through isolation amplifiers AD210AN, with $10^{12}$ Ω input and <1 Ω output impedance. The isolation amplifiers are employed here mainly to provide impedance matching between dividers and the data-acquisition system. The heating voltage can be deduced from measured floating potentials and the heating current, $I_H$, is measured using a 0.05 Ω current shunt and Encore Electronics isolation amplifier FL644-002 with gain set to 5.

## IV. EXPERIMENTAL PROCEDURE

To test if radial probe insertion disturbs the near-anode plasma, yet another, floating cylindrical probe (with 0.25 mm wire diameter and 3 mm uncovered tip length) was introduced into the channel at various thruster operating conditions, namely discharge voltages, $V_d$, from 200 to 700 volts and mass flow rates, $\dot{m}$, from 2 to 4.9 mg/s. During the first experimental session, the probe was inserted as deep as 20 mm into the channel at several distances from the anode, Z = 2 – 12 mm, and left in plasma for up to 5 min. During the second experimental session, the probe was moved from the anode side of the slot, Z = 2 mm, to the cathode side, Z = 12 mm, at several distances from the thruster axis, R = 37 – 62 mm, i. e. from near the inner wall to near the outer wall. As can be seen from Fig. 3, the discharge current vs. time characteristics indicate that the motion of the probe near the anode does not cause perturbations to the Hall thruster discharge. This leads to the conclusion that radial probe insertion is suitable for measurements in the near-anode region of a Hall thruster.

The biased and emissive probes were then used to characterize near-anode processes in Hall thrusters. The biased flat probe data was acquired at the rate of 2000 samples per second for 6 seconds, in order to provide 12 sweeps of the biasing voltage (6 up and 6 down) for each measurement. A biased probe voltage-current characteristic for $V_d = 450 V$ and $\dot{m} = 3\,mg/s$, smoothed over 40 points by using FFT filter and corrected for DC-offsets of isolation amplifiers and voltage drop over the current shunt, is given in Fig. 4.a. The measured discharge current is 2.94 A, the plasma potential relative to the anode is –6.8 V, the electron temperature is 3.9 eV and the plasma density is $2.5 \cdot 10^{11}\,cm^{-3}$. The plasma potential and the electron saturation current were determined graphically by finding the knee in the electron part of a probe voltage-current characteristic. The electron temperature was determined by plotting $\ln(I_{PR})$ vs. $V_{PR}$ and finding a slope of the straight part of the curve. The plasma density was then deduced from the electron saturation current and the electron temperature assuming a Maxwellian electron distribution function (EDF). The repeatability of the biased probe measurements was estimated graphically by plotting all probe V-I characteristics acquired in one data set on the same graph. The spatial resolution of the biased probe in the axial direction was estimated to be the probe diameter, 0.76 mm. The effects of the magnetic field and the flowing plasma are believed to be insignificant for measurements in the near-anode region of a Hall thruster.

The emissive probe data was acquired at the rate of 6000 samples per second for 1 second, in order to provide 60 off half-cycles for each measurement. Fig. 4.b shows the emissive probe floating potential vs. time characteristic, $\varphi_{fl}^{em}(t)$, for $V_d = 450\,V$ and $\dot{m} = 3\,mg/s$, corrected for the DC-offset of the isolation amplifier. The measured discharge current is 2.84 A, the plasma potential relative to the anode is –5.5 V and the electron temperature is 4.6 eV. The probe floating potential saturated when the amplitude of the heating current was greater than 6 A, which corresponds to the heating power of approximately 12 Watts. The floating potential of the hot probe was averaged over 25 data points near the middle of the off half-cycle (to ensure zero heating current), and then averaged over 60 off half-cycles. The cold probe data was averaged over all 6000 points to find a floating potential. The cathode potential and the discharge voltage were measured simultaneously with the probe floating potential in order to deduce the probe potential relative to the anode.

Due to a potential drop in the sheath and pre-sheath formed between the probe surface and the plasma, saturated floating potential of the emissive probe appears to be smaller than the plasma potential. Following Schwager,[14] we can estimate the true plasma potential, $\varphi_{pl}$, for Xenon plasma as: $\varphi_{pl} \approx \varphi_{fl}^{em} + 1.5 \cdot T_e/e$, where $T_e$ is the electron temperature. Assuming Maxwellian EDF we can deduce $T_e$ using a classical expression for the difference between the plasma potential and the floating potential of a cold probe, $\varphi_{fl}^{cl}$, which for Xenon plasma becomes: $\varphi_{fl}^{cl} = \varphi_{pl} - 5.77 \cdot T_e/e$. The final expression for $\varphi_{pl}$ using measured $\varphi_{fl}^{em}$ and $\varphi_{fl}^{cl}$ can be then obtained as:

$$\varphi_{pl} \approx 1.35 \cdot \varphi_{fl}^{em} - 0.35 \cdot \varphi_{fl}^{cl}.$$

The standard deviation of the measured plasma potential was estimated as $\sigma^{pl} = 1.35 \cdot \sigma^{em} + 0.35 \cdot \sigma^{cl}$, where $\sigma^{em}$ and $\sigma^{cl}$ are the standard deviations of the floating potentials measured with the hot and the cold probes, respectively. The assumption of the Maxwellian EDF introduces an additional uncertainty in determination of the electron temperature (and therefore the plasma potential), which is hard to estimate.

## V. SUMMARY AND CONCLUSIONS

Experimental results show that radial probe insertion does not cause perturbations to the discharge and therefore can be used for accurate near-anode measurements. Through probing of the 2kW Hall-thruster plasma in several radial and axial locations, near-anode plasma structure and anode sheath were successfully characterized.[15] The experimental data obtained from biased and emissive probes clearly indicates that the near-anode plasma structure is essentially two-dimensional, and that the anode sheath strongly depends on the Hall thruster operating conditions. The results of the measurements are currently being analyzed to provide a more complete picture of near-anode processes in a Hall discharge.

## ACKNOWLEDGEMENT

The authors would like to thank Mr. D. Staack for his contribution to preparations and delivering of the experiments. We also benefited from discussions with Dr. V. Semenov and Mr. A. Smirnov. The authors are indebted to Mr. G. D'Amico for excellent technical support. This work was supported by the US DOE under contract No. DE-AC02-76CH03073.

APPENDIX: *CORRECTION TO THE FLOATING POTENTIAL OF THE EMISSIVE PROBE*

It was shown in Ref. 16 that the increase of the emissive probe temperature does not necessarily lead to saturation of the probe floating potential. When the ratio of the emitted current density to the collected current density, $\delta = j_{em}/j_{coll}$, becomes higher than the critical value, $\delta_{cr} \sim 1$, the potential well is formed near the probe surface to limit the flux of emitted electrons (Fig. 5. a). To estimate $\varphi_{pl} - \varphi_{fl}^{em}$, we assume that $\varphi_{pl} - \varphi_{min} \approx 1.5\, T_e/e$,[14] and that the current density of emitted electrons that reach the plasma is $j_{em}^* = \delta_{cr} \cdot j_{coll} \sim j_{coll}$. Assuming Maxwellian EDF for both emitted and collected electrons we can write: $j_{coll} = j_e^{sat} \cdot \exp[e \cdot (\varphi_{pl} - \varphi_{min})/T_e]$ and $j_{em}^* = j_{em} \cdot \exp[e \cdot \Delta\varphi^*/T_{pr}]$, where $j_e^{sat}$ is the electron saturation current density, $e$ is the electron charge, $\Delta\varphi^* = \varphi_{fl}^{em} - \varphi_{min}$ and $T_{pr}$ is the probe temperature. Thus, we obtain: $\varphi_{pl} - \varphi_{fl}^{em} = 1.5\, T_e/e - \Delta\varphi^*$ and $\Delta\varphi^* \sim \ln(4.5 \cdot j_{em}/j_e^{sat}) \cdot T_{pr}/e$. Using the Richardson-Dushman's formula, $j_{em} = A \cdot T_{pr}^2 \cdot \exp[e \cdot \varphi_w/T_{pr}]$, where $A = 3\, A/(cm^2 \cdot K^2)$ and $\varphi_w = 2.63\, eV$ for thoriated tungsten,[16] and $j_e^{sat} = 0.4\, A/cm^2$, which is typical for the near-anode plasma in the 2 kW Hall thruster, we can finally deduce: $\Delta\varphi^* \sim \ln(5.8 \cdot T_{pr}) \cdot T_{pr}/5800 - 2.63$, where $T_{pr}$ is in $K^o$. As can be seen from Fig. 5. b, $\Delta\varphi^*(T_{pr})$ can be approximated with a linear function: $\Delta\varphi^* \sim T_{pr}/560 - 2.97$. For $T_{pr} = 2000 K$ and $T_e = 5\, eV$ we have $\Delta\varphi^*/(1.5\, T_e/e) = 8\, \%$.

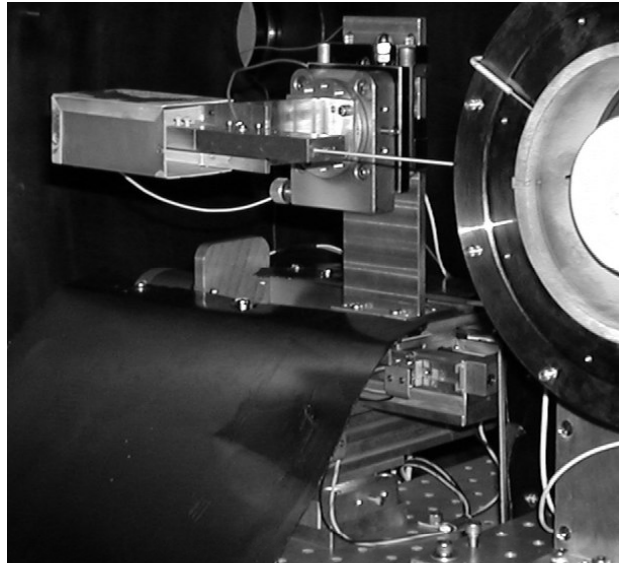

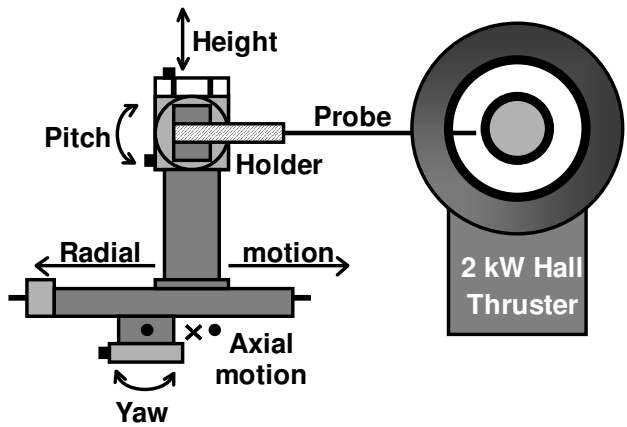

Fig. 1. Radial probes diagnostic setup for near-anode measurements.

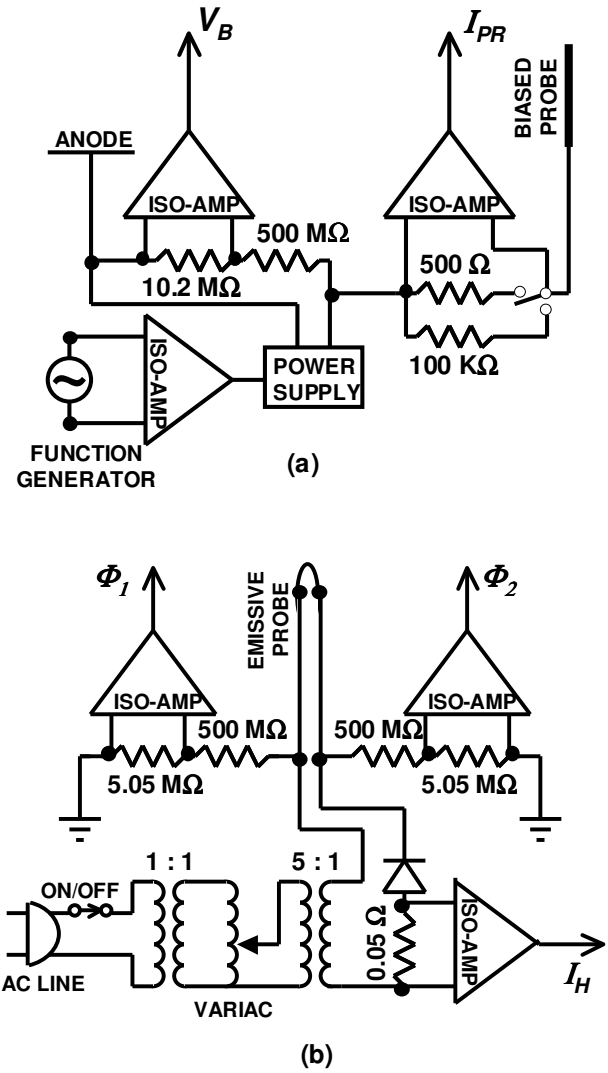

Fig. 2. Electronic circuit diagrams: (a) Biased probe. (b) Emissive probe

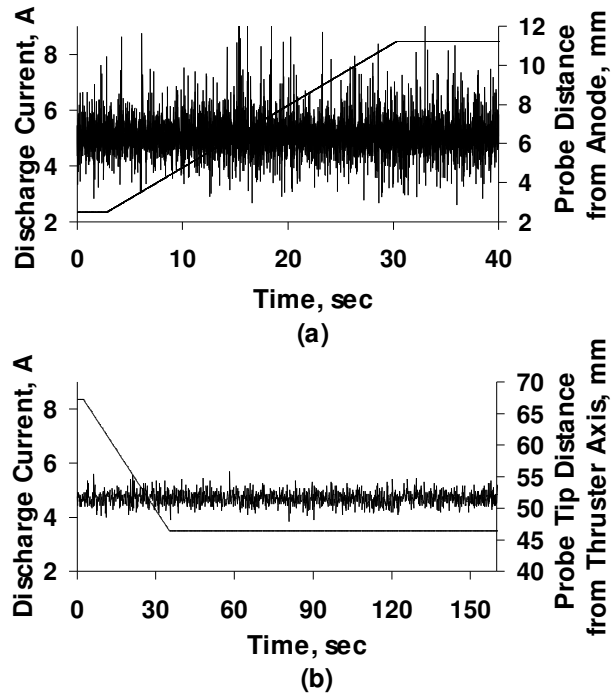

Fig. 3. Discharge current vs. time characteristics for $V_d = 500\,V$ and $\dot{m} = 5\,mg/s$:

a. Radial insertion at Z = 2 mm. Sampling rate = 10 samples/sec.

b. Axial motion at R = 55 mm. Sampling rate = 200 samples/sec.

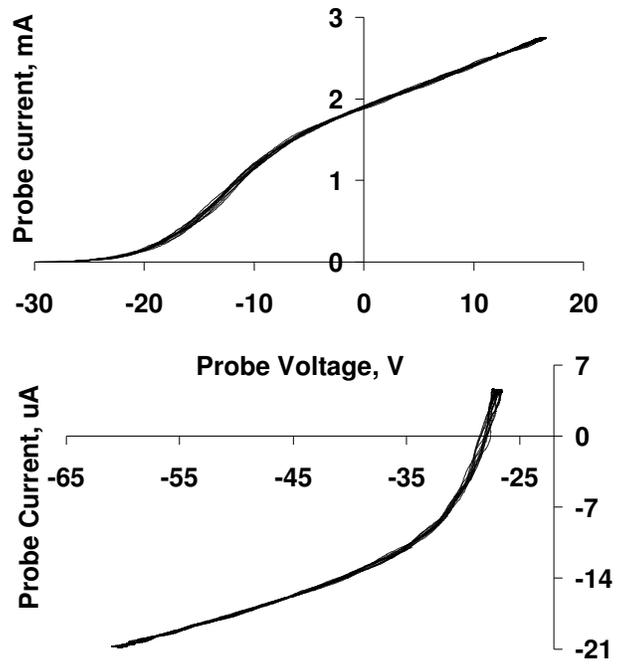

(a)

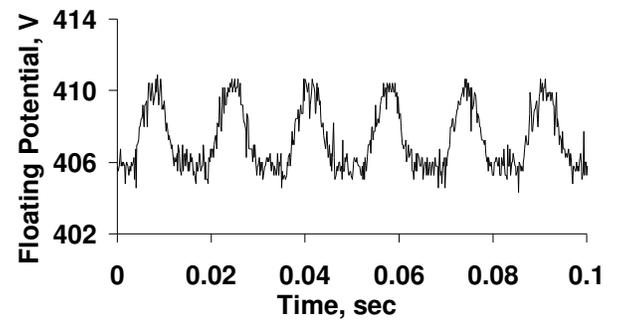

(b)

Fig. 4. Experimental data for $V_d = 450\,V$ and $\dot{m} = 3\,mg/s$.

At the middle of a slot, Z = 7mm, and channel median, R = 49 mm:

a. Electron and ion parts of the biased probe voltage-current characteristic.

b. Emissive probe floating potential vs. time characteristic, $\Phi_2(t)$. $I_H = 6.26A$.

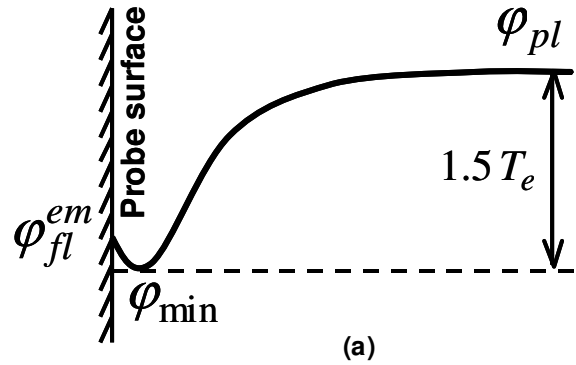

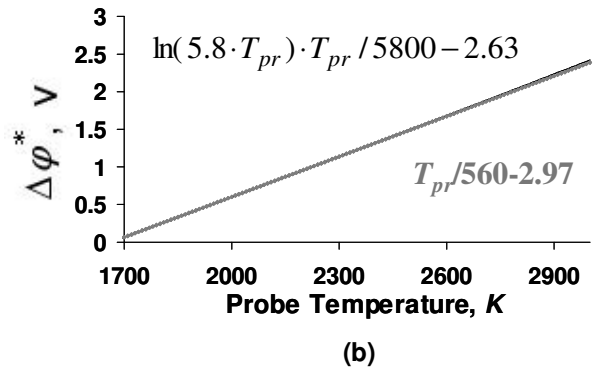

Fig. 5. Correction to the floating potential of the emissive probe:

a. Shape of the near-probe potential profile for $\delta > \delta_{cr}$.

b. Graph of $\Delta\varphi^*(T_{pr})$ with linear approximation for $j_e^{sat} = 0.4\ A/cm^2$.